\documentclass[aps,prc,twocolumn,showpacs,superscriptaddress,floatfix,10pt]{revtex4-1}
\usepackage{graphicx}% Include figure files
\usepackage{dcolumn}% Align table columns on decimal point
\usepackage{bm}% bold math
\usepackage{amsfonts}
\usepackage{amssymb}
\usepackage{amsmath}

\newcommand{\al}{\alpha}

\newcommand{\az}{\varphi}
\newcommand{\ro}{\rho}
\newcommand{\om}{\omega}

\newcommand{\la}{\lambda}
\newcommand{\be}{\beta}

\newcommand{\oeq}{\begin{equation}}
\newcommand{\ceq}{\end{equation}}
\newcommand{\oeqn}{\begin{eqnarray}}
\newcommand{\ceqn}{\end{eqnarray}}

\renewcommand{\>}{\rangle}
\newcommand{\<}{\langle}
\renewcommand{\(}{\left(}
\renewcommand{\)}{\right)}

\renewcommand{\ll}{\left|}
\newcommand{\rl}{\right|}

\newcommand{\stf}{\,\,\,}
\newcommand{\sdf}{\,\,}
\newcommand{\stb}{\!\!\!}

\newcommand{\oQ}{\hat{Q}}

\newcommand{\oH}{\hat{H}}

\newcommand{\oV}{\hat{V}}
\newcommand{\oU}{\hat{U}}
\newcommand{\oD}{\hat{D}}
\newcommand{\oR}{\hat{R}}

\newcommand{\oh}{\hat{h}}

\newcommand{\ovr}{\hat{\bf r}}

\newcommand{\op}{\hat{p}}

\newcommand{\oad}{\hat{a}^\dagger}
\newcommand{\oa}{\hat{a}}

\renewcommand{\d}{{\mbox d}}

\newcommand{\hb}{\hbar}

\renewcommand{\vr}{{\bf r}}

\newcommand{\vj}{{\bf j}}
\newcommand{\vT}{{\bf T}}
\newcommand{\vJ}{{\bf J}}

\newcommand{\vR}{{\bf R}}

\newcommand{\vS}{{\bf S}}

\newcommand{\vp}{{\bf p}}

\newcommand{\mH}{{\mathcal{H}}}

\begin{document}

%\title{From microscopic bare nucleus-nucleus potential to coupled-channels effects on fusion}
\title{Microscopic approach to coupled-channels effects on fusion}

\author{C. Simenel}
\author{M. Dasgupta}
\author{D. J. Hinde}
\author{E. Williams}
\affiliation{Department of Nuclear Physics, RSPE, Australian National University, Canberra, ACT 0200, Australia}

\date{\today}

%------------------------------------------------------------------------------
\begin{abstract}
\begin{description}
\item[Background]  Near-barrier fusion can be strongly affected by the coupling between relative motion and internal degrees of freedom of the collision partners.
The time-dependent Hartree-Fock (TDHF) theory and the coupled-channels (CC) method are standard approaches to investigate this aspect of fusion dynamics.
However, both approaches present limitations, such as a lack of tunnelling of the many-body wave function in the former and a need for external parameters to describe the nucleus-nucleus potential and the couplings in the latter. 
\item[Methods]     A method combining both approaches is proposed to overcome these limitations. 
CC calculations are performed using two types of inputs from Hartree-Fock (HF) theory: the nucleus-nucleus potential calculated with the frozen HF method, and the properties of low-lying vibrational states and giant resonances computed from the TDHF linear response. 
\item[Results]     The effect of the couplings to vibrational modes is studied in the $^{40}$Ca$+^{40}$Ca and $^{56}$Ni$+^{56}$Ni systems.  
This work demonstrates that the main effect of these couplings is a lowering of the barrier, in good agreement with the fusion thresholds predicted by TDHF calculations. 
\item[Conclusions] 
As the only phenomenological inputs are the choice of the internal states of the nuclei and the parameters of the energy density functional used in the HF and TDHF calculations, 
the method presented in this work has a broad range of possible applications, including studies of alternative couplings or reactions involving exotic nuclei.
\end{description}
\end{abstract}
\pacs{25.70.Jj,24.10.Eq,21.60.Jz}% PACS, the Physics and Astronomy Classification Scheme.
\maketitle

%------------------------------------------------------------------------------
\section{Introduction}

% ETW reworked intro - Cedric, I have been lazy on references, as I am on mobile internet %

Our understanding of nuclear reactions is shaped by two extreme perspectives: a macroscopic picture and a microscopic picture.  In the former, the colliding atomic nuclei are treated as charged liquid drops undergoing a large-scale shape evolution. In the latter, the internal structure of the nuclei at the nucleon level plays a role in the final outcome of the reaction. The two perspectives are connected: the macroscopic evolution of the colliding system must be directly related to the microscopic behavior of its constituent nucleons.  However, the complexity of the nuclear many-body problem typically requires a compromise between the two perspectives when it comes to producing models capable of shedding light on our observations of the sub-atomic world. 

Nowhere is this compromise more evident than in models of heavy ion collisions near the Coulomb barrier. In such collisions, a wide range of reaction mechanisms---including inelastic scattering, (multi-)nucleon transfer, fusion, quasifission, and compound nucleus fission---can all compete with each other. The competition between all of these outcomes can be profoundly shaped by both the large-scale evolution of the two-nucleus system and the internal structure of the nuclei involved. 

One of the clearest examples of this dual influence can be found in the coupling between the internal degrees of freedom of the colliding nuclei and their relative motion. Through coupling, these internal degrees of freedom, which can include low-lying vibrations \cite{mor94,ste95a,das98,bal98}, rotations \cite{lei95}, and higher-lying giant resonances \cite{new04,dia08}, have been shown to modify dynamically the interaction potential between the collision partners and, subsequently, the outcome of the reaction itself \cite{das85}. 

Signs of this coupling can be extracted from experimental fusion cross sections $\sigma_{fus}$ for reactions around the Coulomb barrier. By taking the second derivative of $E\sigma_{fus}$ to derive a quantity known as the experimental barrier distribution, a direct visualization of the effect of the couplings can be---and in many cases, has been \cite{das98}---obtained \cite{row91}.

The standard theoretical approach to study these couplings between internal degrees of freedom and relative motion is the coupled-channels method. 
The method is macroscopic in its approach, and includes three ingredients: a collective rotational or vibrational model of nuclear structure for the target and projectile, an interaction potential between the nuclei, and in some cases, a simple transfer potential.  
Like most macroscopic models, these potentials and the characteristics of the internal degrees of freedom affecting the relative motion are provided as input parameters. 
Coupled channel codes are then not predictive on their own as they require input parameters from either experimental data or from theoretical calculations.

%% I haven't really modified the stuff below. It is too detailed (and contributes to your long intro) -- what do you think the most essential points are, in following from the reworked intro above? Anything else can potentially be moved into a more detailed discussion of TDHF in a Theory section below. - ETW

The coupling between relative motion and internal degrees of freedom can in principle be addressed through a different theoretical approach, investigating the path to fusion through {\it microscopic} models with effective interactions between the nucleons. The time-dependent Hartree-Fock (TDHF) theory is one such approach, and has been widely used to study reaction dynamics near the barrier (see Ref. \cite{sim12b} for a recent review). 

One difficulty of these dynamical microscopic studies is to disentangle the role of various internal degrees of freedom on the fusion process. 
Indeed, in the TDHF approach, all the couplings are automatically included at the mean-field level and it may be difficult, for instance, to separate the contribution of transfer from that of vibrational states. 
In addition, one cannot extract directly realistic barrier distributions from standard TDHF calculations as the quantum tunneling of the many-body wave-function, where the coupling effects are most apparent, is not accounted for \cite{neg82}.  

It would be highly desirable to combine the advantages of the coupled-channels model with those of the TDHF theory. 
Ideally, one would like to be able to predict realistic fusion cross-sections (and thus the experimental barrier distributions) using inputs from microscopic models alone, removing the need to rely on quantities extracted from experimental data. % Modified this last sentence a bit - ETW %

% I would rework the section below to emphasize the approach as a proof of principle for bridging the microscopic and macroscopic approaches, as a first step towards achieving a fully microscopic model of nuclear reactions.  Only include the essential ingredients here; details can come later. - ETW %
% Emphasize the fact that CCFULL is not predictive on its own as it requires input parameters from either exp data or from theoretical calculations.

As a first step towards achieving a fully microscopic model of nuclear reactions, we propose a method based on the coupled-channel framework where only the effective interaction between the nucleons is required as an input.
More precisely, apart from the choice of the number of phonons of each mode, the parameters of the coupled-channel calculations are entirely determined from (time-dependent) Hartree-Fock calculations.

As a proof of principle, we investigate the effect on the centroid of the barrier distribution of the coupling between vibrational states and the relative motion.
For this purpose, we consider light symmetric doubly magic systems to minimise the role of transfer and coupling to rotational states. 
In particular, the $^{40}$Ca+$^{40}$Ca and $^{56}$Ni+$^{56}$Ni systems are studied to compare the role of magic numbers 20 and 28 on the vibrational couplings.

We start with a general discussion on the TDHF and coupled-channels models as well as a presentation of the method combining both approaches in section \ref{sec:formalism}. 
We then provide details of the Skyrme HF formalism in section \ref{sec:SHF}. 
The technique used to compute the bare nucleus-nucleus potential is introduced in section \ref{sec:frozen}. 
Near-barrier TDHF calculations are presented in section \ref{sec:TDHF}. 
The strength functions of vibrational modes are computed in section \ref{sec:RPA}, 
and are used to get the energy and deformation parameters of the main vibrational modes in section \ref{sec:param}.
The coupled-channels calculations are performed in section \ref{sec:cc}. 
Finally, we discuss the differences between $^{40}$Ca+$^{40}$Ca and $^{56}$Ni+$^{56}$Ni systems in section \ref{sec:nini}.

\section{Combining microscopic and macroscopic formalisms \label{sec:formalism}}

This section is devoted to a general discussion of the theoretical approaches.
The applications of the TDHF method to both nuclear vibrations and heavy-ion collisions is first discussed.
Then the Coupled-Channel approach to determine the effect of couplings to internal degrees of freedom on fusion is introduced. 
Both approaches have limitations in their predictive power for fusion reactions. 
We aim at overcoming these limitations in combining them as described at the end of this section.

\subsection{Successes and limitations of the TDHF approach to describe nuclear dynamics}

A natural application of the TDHF formalism is to investigate nuclear vibrations at the mean-field level using the linear response theory  \cite{sim12b}.
Indeed, the linearization of the TDHF equation leads to the random-phase approximation (RPA) \cite{rin80} which is the basic tool to investigate collective vibrations in the harmonic picture. 
Note that, using a TDHF code, the calculations are fully self-consistent, i.e., all terms of the RPA residual interaction are taken into account, including Coulomb and spin-orbit terms.
Most of the numerical applications to investigate nuclear vibrations with time-dependent microscopic models have been focussed on giant resonances~\cite{sim03,ste04,uma05,mar05,rei07,sim09,ste10,fra12,ave13,sca13b}. 
Nevertheless, some applications to the study of collective bound states have also been considered ~\cite{ave08,sim12b,sim13b,sca13b}. 

As mentioned in the introduction, the TDHF approach has also been widely used to study heavy-ion collisions \cite{sim12b}.
However, sub-barrier fusion cannot be described with the TDHF theory because it does not take into account the tunneling of the many-body wave-function. 
Nevertheless, fusion thresholds can be computed as the energy above which central collisions lead to fusion while the exit channel at lower energies is made of two outgoing fragments.
As a result, realistic three-dimensional TDHF calculations of fusion thresholds are in excellent agreement with the centroids of experimental barrier distributions \cite{sim08,was08}. 
In addition, the role of deformation and reorientation on fusion \cite{sim04,uma06c}, as well as the effect of transfer channels \cite{sim08,uma08a,was09c}, have also been studied with TDHF calculations. 
Note that both Coulomb and nuclear contributions are accounted for self-consistently and that no weak coupling assumption \cite{hag97b} is made, i.e., the couplings are treated to all orders. 

The ability of TDHF to describe collective vibrations at the mean-field level implies that it can also be used to investigate the interplay between vibrations and reaction dynamics. 
For instance, collective vibrations of the non-equilibrated system after capture have been studied \cite{sim01,sim07,iwa10a,obe12}. 
Low-lying collective vibrations have also been shown to be excited in a recent TDHF application to $^{16}$O+$^{16}$O collisions just below the barrier \cite{sim13b}. 
In these calculations, the octupole moment of the outgoing fragments was oscillating and these oscillations could be associated with the excitation of the $3^-_1$ state in $^{16}$O.

The above discussion shows that the TDHF framework can be used to investigate vibrational states as well as their coupling to the relative motion.
Associated with a modern coupled-channel code such as \textsc{ccfull}, it is then well-equipped to provide realistic fusion cross-sections around the fusion barrier. 

\subsection{Coupled-channel framework}

In the coupled-channels model, the relative motion between the collision partners is affected by a potential generated by the Coulomb and nuclear interaction between the nuclei.
A standard form of the nuclear part of the nucleus-nucleus potential is given by the Woods-Saxon function
\oeq
V_{WS}(D)=\frac{-V_0}{1+\exp\frac{D-r_0(A_1^{1/3}+A_2^{1/3})}{a}},
\label{eq:WS}
\ceq
where $V_0$ is the depth of the potential and $a$ its diffuseness.

The excitation of vibrational states induces a variation of the distance between the surface of the nuclei.
It can be accounted for by replacing the radii of the collision partners $R_{0_i}=r_0A_i^{1/3}$ in Eq.~(\ref{eq:WS}) by the observable $\oR_i(\theta\phi)$ measuring the distance to the surface of the nucleus $i$ defined, e.g., as the isodensity with $\ro_0/2$, where $\ro_0=0.016$~fm$^{-3}$ is the saturation density. 
It could be written as (see, e.g., Ref.~\cite{hag12})
\oeq
\oR(\theta\phi)\simeq R_0+\sum_{\lambda\ge2}  \sum_\nu \frac{R_0\be_{\lambda}^{(\nu)}}{\sqrt{2\la+1}} \(\hat{a}^{\dagger(\nu)}_{\la\mu}+ (-)^\mu\oa_{\la\mu}^{(\nu)}\) Y_{\la\mu}^*(\theta\phi),
\label{eq:oR}
\ceq
where %$\obe_\la=\sum_\nu\be_{\la}^{(\nu)}|\nu\>\<\nu|$ and 
$\be_{\la}^{(\nu)}$ is the deformation parameter of the phonon $|\nu\>$. 
The operators  $\hat{a}^{\dagger(\nu)}_{\la\mu}$ and $\oa_{\la\mu}^{(\nu)}$ create and annihilate, respectively, a phonon $|\nu\>$ with angular momentum $\la$ and projection $\mu$. 
In the isocentrifugal approximation, the spherical
 harmonics disappears in Eq.~(\ref{eq:oR}) and only the $\mu=0$ component remains \cite{hag12}.
 The potential including all order couplings is then obtained by transforming the Woods-Saxon potential coordinate according to 
\oeq
\oV\equiv V_{WS}\(\oD-\sum_{i=1}^2 R_{0_i}\sum_{\nu,\la\ge2} \frac{\be_{\la_i}^{(\nu)}}{\sqrt{4\pi}}\(\hat{a}^{\dagger(\nu)}_{\la0}(i)+\oa_{\la0}^{(\nu)}(i)\)\).\nonumber
\ceq

In principle, the above technique can be used to incorporate the effect of both collective low-lying vibrations and giant resonances.
In particular, low-lying energy collective vibrations effectively lead to a fragmentation of the single-barrier generating a barrier distribution \cite{mor94,ste95a,das98,bal98}. 
Collective states at higher energies, like giant resonances, can also affect the potential barrier. 
However, they essentially induce a global shift of the barrier distribution without modifying its shape \cite{hag97a}. 
The same effect is obtained in the case of light systems as the small product of the proton numbers $Z_1Z_2$ leads to a small coupling strength. 
For instance, the coupling to the $3^-_1$ state of $^{40}$Ca with light collision partners or of $^{16}$O with any target essentially produces an adiabatic renormalization of the static potential without changing the shape of the barrier distribution \cite{das98}.

In standard applications of the coupled-channels method, e.g., with the code \textsc{ccfull} \cite{hag12}, the collective model (using energies and deformation parameters extracted from experiment) is used to investigate the {\it shape} of the experimental barrier distribution, while the parameters of the nucleus-nucleus potential are adjusted to reproduce its {\it centroid}. This method has been quite successful in explorations of reactions on target nuclei such as isotopes of Sm \cite{lei95} and has been widely applied because of its simplicity. However, it is inherently limited in scope. There are three reasons for this:
\begin{enumerate}
\item The collective picture (a vast simplification of collective nuclear structure in itself \cite{boh75}) requires experimental inputs that are not always available from precision measurements. Although the energy of the states are usually  well known in stable nuclei, this is not always the case for the deformation parameters. As an example, the experimental reduced electric-octupole transition probability $B(E3;0^+_1\rightarrow3^-_1)$ values, from which the deformation parameter $\beta_3$ is computed, varies by more than a factor of two for $^{40}$Ca \cite{kib02}. Obtaining such experimental structure data on exotic nuclei will also be a difficulty limiting the possibility of reaction studies with upcoming radioactive beams. 
\item This approach cannot be used for quantitative prediction of the centroid of the barrier distribution. This is because of the fitting of the barrier centroid, which essentially incorporates the potential renormalization effect due to the higher energy states into the nucleus-nucleus potential.  As a result, the approach can only be used to study of the effect of the couplings to low-energy states; it provides no information on the bare nucleus-nucleus potential.  This limitation is a significant one: a key difficulty in predicting the effect on the barrier centroid of the couplings to vibrational states has been due to a poor knowledge of the bare nucleus-nucleus potential \cite{gon04}.
\item 
%The contribution of transfer to the experimental barrier distribution is currently poorly understood. %Cedric, I admit that I have not played with the transfer component of CCFULL enough to say much about this off the top of my head. I want to say that we don't have a good way of separating out the contribution of transfer from that of structure, but this point will have to be strengthened. Think of this as a placeholder / reminder to modify the above to at least allude to transfer.
%
There are indications that (multi-)nucleon transfer channels play an important role in the fusion process \cite{ste95b,jia10,eve11,mon13}.
However, despite progress \cite{pol13}, a proper treatment of the interplay between transfer channels and fusion is still lacking so far.
\end{enumerate}
From these limitations, it is clear that a consistent approach to compute both $(i)$ the bare nucleus-nucleus potential and $(ii)$ the energy and deformation parameter of the states is required. 

\subsection{Description of the proposed method}

%% Consider moving the following (up until %%END%%) to a later section - it's a lot of detail to include in the intro - ETW %%

In this work, a method is proposed to describe the effect on the fusion process of the coupling to {\it vibrational} states where the only input is the Skyrme effective interaction \cite{sky56}. 
It should be noted that the fit protocol of the parameters of this interaction does not involve input from reaction mechanisms such as cross-sections or Coulomb barriers (see, e.g., Ref. \cite{cha98}). 
The basic steps of the approach are:
\begin{enumerate} 
\item The bare nucleus-nucleus potential is computed from the frozen Hartree-Fock (HF) technique. 
\item Near-barrier TDHF calculations are used to determine the fusion threshold including dynamical effects at the mean-field level. 
\item The same TDHF code is used to compute the strength function of these modes using the linear response theory.
\item The strength function is used to extract the energy and deformation parameter of collective vibrational states.
\item The bare nucleus-nucleus potential and the parameters of the coupling are used in standard coupled-channels calculations to determine fusion cross-sections. 
\end{enumerate}

Note that step 2 is not really mandatory to compute the final fusion probabilities. 
However it provides a benchmark and, if the centroid of the final barrier distribution is in good agreement with the TDHF fusion threshold, we can reasonably conclude that the most relevant internal degrees of freedom have been included in the coupled-channels calculations. 
%% END %%
%

%Our primary goal in this work is then to investigate the role of the different vibrational couplings on the centroid of the barrier distribution. 
%In particular, the effect on the potential renormalization of the coupling to collective low-lying states and to states at higher energies such as giant resonances is studied. 

In order to reach this goal, we will focus on the light symmetric doubly magic systems. 
This choice is motivated by the following reasons:
\begin{itemize}
\item Nucleon transfer is not favored in symmetric systems.
\item Doubly magic nuclei have no pairing, hence the TDHF description in terms of independent particle states should be sufficient.
\item The colliding partners are spherical, implying that no coupling to rotational states is to be expected.
\item The $Z_1Z_2$ product is small in light systems. This results in a smaller coupling strength and the main effect of the couplings is an adiabatic renormalization of the potential \cite{hag97a}, i.e., a  shift of the barrier which simplifies the comparison with the TDHF fusion threshold. 
\end{itemize}

\section{The Skyrme Hartree-Fock microscopic framework \label{sec:SHF}}

Vautherin and Brink made an important breakthrough in the 1970's when they performed the first Skyrme Hartree-Fock calculations of atomic nuclei \cite{vau72}. 
With a small number of parameters, mean-field and beyond mean-field calculations based on the Skyrme energy-density-functional \cite{sky56} with pairing residual interaction provided a good description of binding energies and static deformations across the nuclear chart \cite{ben06}. 

Pairing correlations play an important role in the description of nuclear ground-states \cite{boh58}. 
These correlations have been recently included in microscopic time-dependent approaches as well \cite{ave08,eba10,ste11,has12,sca12}.
However, to simplify the presentation of the method, the case of doubly magic nuclei is considered here. 
Indeed, the latter exhibit no pairing correlations at the mean-field level and can be described within the Hartree-Fock theory. 
The more general case of deformed nuclei including pairing correlations will be the subject of future work. 

\subsection{The HF and TDHF formalisms}

The static and time-dependent Hartree-Fock approaches to the nuclear many-body problem have been discussed in many works (see, e.g., the text book by Ring and Schuck \cite{rin80}). 
Here, we briefly summarize the main aspects of the theory. 

The mean-field description of many interacting particles introduced by Hartree \cite{har28} has been extended by Fock to properly take into account the Pauli principle in the case of identical fermions \cite{foc30}. 
In this case, the independent-particle state is written as a Slater determinant of single-particle wave functions. 
All the information on the system is then contained in the set of occupied single-particle states $|\az_{i=1\cdots A}\>$, or, equivalently, in the one-body density-matrix with elements $\ro_{\al\be}=\sum_{i=1}^A\<\al|\az_i\>\<\az_i|\be\>$, where $|\al\>$ and $|\be\>$ are single-particle states, and  $A$ is the total  number of particles.

In the Hartree-Fock approach, each particle evolves independently in the self-consistent mean-field $\oU[\ro]$ generated by the other particles. 
The Hartree-Fock ground-state is obtained from the $A$ eigenstates of $\oh[\ro]=\frac{\op^2}{2m}+\oU[\ro]$ with the lowest eigenvalues $e_i$, i.e., obeying $\oh[\ro]|\az_i\>=e_i|\az_i\>$.  

In the original formulation of the HF theory, the elements of the matrix representing the single-particle Hamiltonian $\oh[\ro]$ are computed from the many-body Hamiltonian $\oH$ according to 
\oeq
h_{\al\be}[\ro]=\<\al|\oh[\ro]|\be\>=\frac{\delta\<\Phi|\oH|\Phi\>}{\delta \ro_{\be\al}},
\label{eq:h}
\ceq
where $|\Phi\>$ is the independent-particle state of the system.

The time-dependent extension of the HF theory was proposed by Dirac in 1930 \cite{dir30}. 
In this approach, the occupied single-particle states obey the Schr\"odinger-like equations
$i\hb\frac{d}{dt}|\az_i(t)\> = \oh[\ro] |\az_i(t)\>$ which can be expressed using the $\ro$ and $h$ matrices as
\begin{equation}
i\hbar \frac{d\rho}{dt} =\left[h[\rho],\rho\right].
\label{eq:TDHF}
\end{equation}
Eq.~(\ref{eq:TDHF}) is the TDHF equation. 
Its static limit, i.e., setting the left-hand side to zero, is the HF equation. 

The TDHF equation is non-linear due to the self-consistency of the mean-field potential. 
This allows for a proper treatment of collective phenomena, such as vibrations. 
This is also crucial for the inclusion of one-body dissipation mechanisms, for example resulting from the collisions of the nucleons on the mean-field wall at the surface. 

\subsection{The Skyrme energy-density functional \label{sec:skyrme}}

In nuclear physics, the average energy $\<\Phi|\oH|\Phi\>$ in Eq.~(\ref{eq:h}) is usually replaced by an energy density functional (EDF) $E[\ro]$ derived from  Skyrme \cite{sky56} or Gogny \cite{dec80} phenomenological effective interactions and containing the Coulomb repulsion between protons. 
Effective interactions are used instead of the bare interaction {\it(i)} to avoid divergences of the mean-field due to the hard-core repulsion at short distances and {\it(ii)} to sum up some many-body effects (see chapter~4 of Ref.~\cite{rin80} for more details). 

The main difference between the Skyrme and Gogny interactions is their range: the Skyrme interaction is a contact (zero-range) interaction, while the Gogny one has a finite range. 
Due to its zero-range nature, it is easier to use the Skyrme interaction on cartesian grids. 
For this reason, almost all TDHF calculations have been done using the Skyrme EDF (see Ref.~\cite{has12} for recent TDHF calculations of vibrations with the Gogny interaction).

In general, the EDF is a function of the entire one-body density-matrix including non-local terms $\rho(\vr s q, \vr' s' q')$ where $s$ and $q$ are the spin and isospin of the nucleons. 
In the Skyrme case, however, the zero-range nature of the interaction implies that the EDF is a functional of local densities only. 
The Skyrme EDF is then expressed as $E[\ro_q,\tau_q,\vj_q,\vJ_q,\vS_q,\cdots]$, where $\rho_q(\vr)$ are the proton ($q=-\frac{1}{2}$) and neutron ($q=+\frac{1}{2}$) densities, $\tau_q(\vr)$ are the kinetic energy densities, $\vj_q(\vr)$ are the current densities, $\vJ_q(\vr)$ are the spin-orbit densities, and $\vS_q(\vr)$ are the spin densities (see, e.g., Ref.~\cite{sim13a} for  explicit expressions of these densities as well as of the Skyrme functional and mean-field). 

For time-reversal invariant systems, e.g., HF ground-states of even-even nuclei, the time-odd densities $\vj_q$ and $\vS_q$ vanish. 
However, they need to be included in time-dependent calculations of heavy-ion collisions to ensure Galilean invariance \cite{eng75}. 
In particular, they provide an important contribution to the one-body energy dissipation \cite{mar06}. 
The spin-orbit interaction also plays a crucial role in the dissipation mechanisms \cite{uma86}. 
This is true even for light systems for which the spin-orbit energy is almost zero in the ground states of the collision partners (e.g., magic numbers up to 20 are reproduced without spin-orbit interaction). 

The most general expression of the Skyrme EDF contains other densities than the ones described above, such as the spin-current pseudo-tensor $\stackrel{\leftrightarrow}{J}$ (only the anti-symmetric part of $\stackrel{\leftrightarrow}{J}$, which is the spin-orbit density $\vJ$, is  included) and the spin-kinetic energy density $\vT$.
 These densities have not always been included in the fit of the EDF parameters as they are expected to provide only small corrections to the energy \cite{cha98}.
The contribution of these additional densities in the dynamics of heavy-ion collisions has been recently tested by Loebl and collaborators~\cite{loe12}.
Although they have been shown to increase dissipation at high energy, their effect in non-violent reactions such as heavy-ion collisions around the Coulomb barrier can be neglected. 
The SLy4$d$ parametrization \cite{kim97} of the Skyrme EDF, which is used in this work and which has been obtained without these additional densities, is therefore expected to provide a very reliable mean-field description of the reaction dynamics at low energies \cite{sim12b}. 

\subsection{Numerical aspects \label{sec:num}}

The increase of computational power has allowed the development of three-dimensional HF and TDHF codes using modern Skyrme functional including spin-orbit interaction. Realistic TDHF calculations of nuclear vibrations \cite{sim03,uma05,mar05,nak05,ste10,fra12} and of heavy-ion collisions \cite{sim01,uma06a,mar06,was08,sek13} have then been made possible.
 Ref.~\cite{sim12b} provides details of the numerical implementation of the TDHF equation.

In this work, the HF ground-states are computed using the \textsc{ev8} code \cite{bon05} without pairing and center of mass corrections. 
The SLy4$d$ parametrization of the Skyrme EDF is used \cite{kim97}. 
The single-particle wave-functions are developed on a cartesian grid with hard boundary conditions.
Using similar numerical approximations, the TDHF equation is solved iteratively in time with a plane of symmetry using the \textsc{tdhf3d} code \cite{kim97}. 

In order to study nuclear vibrations, a time-dependent perturbation $\oV(t)$ is  applied to the HF ground-state (see section~\ref{sec:linear-response}).  
In these calculations, good convergence of the vibrational state properties are obtained with a time step $\Delta t=1.5\times10^{-24}$~s and a grid size $(28\Delta x)^3$  with a mesh grid $\Delta x=0.8$~fm.

At the initial time of the TDHF calculations of heavy-ion collisions, a Galilean boost of the form 
\oeq
|\az_i(t=0)\>=\exp(\frac{i}{\hb}\vp_\al\cdot\ovr)|\az^{HF}_i\>
\ceq
is applied to the  single-particle states $\az^{HF}_i$ of the HF ground-state of the nucleus $\al$, inducing a momentum $\vp_\al$ to its nucleons.
The initial distance between the centers of mass of the nuclei is $D_0=45.6$~fm. 
This value is large enough to include most of the long-range Coulomb excitation which could affect slightly the position of the barrier.

Numerical cartesian grids preserve exactly Galilean invariance only in the limit $\Delta x\rightarrow0$. 
Indeed, finite mesh grids may induce a small spurious excitation of the collision partners. 
In order to minimize this effect while keeping the computational time to a reasonable level, we found that the choice of numerical parameters 
$\Delta x=0.6$~fm and $\Delta t=5\times10^{-25}$~s provided a good compromise, with a spurious excitation energy due to violation of  Galilean invariance less than 0.1~MeV. 
The TDHF calculations for central collisions are performed using a grid size $114\times38\times38\Delta x^3$. 

\section{Bare potential from the frozen Hartree-Fock technique \label{sec:frozen}}

The {\it bare} nucleus-nucleus potential is defined as the interaction potential between the nuclei in their ground states. 
At an energy well above the barrier, the two nuclei do not have time to rearrange their density before they overcome the barrier and this potential describes properly the interaction between the reactants. 
At an energy closer to the barrier, however, the fusion process is slower and the density has time to encounter rearrangements induced by the couplings to internal degrees of freedom. 
This variation of the density can also induce a change of the potential. 
As a result, the latter can exhibit an energy dependence due to the couplings.

Several techniques have been recently introduced  to compute nucleus-nucleus potentials from dynamical microscopic approaches \cite{uma06b,was08,zhu13}. 
In particular, the density-constrained (DC-TDHF) \cite{uma06b} and dissipative-dynamics (DD-TDHF) \cite{was08} approaches have been developed to extract the nucleus-nucleus potential from TDHF trajectories and to determine its energy dependence. 
%This energy dependence comes from the fact that, at high energy, the fragments do not have time to rearrange their density, while close to the barrier they can acquire some deformation. 
The effect of the couplings is then directly included on the potential, but only in average as the potential extracted from a TDHF trajectory  exhibits a single barrier. 
It is also difficult to disentangle the effects of different collective modes as all of them contribute coherently to the TDHF evolution. 
%These models are then well suited to investigate  dynamical effects on the potential. 

Here, we look for a different approach, as our goal is to investigate the effect of the different vibrational states on the barrier distribution. 
Thus, we aim to produce a {\it bare} potential which does not include any dynamical contribution.
In particular, we want to exclude the effect of the coupling between relative motion and internal degrees of freedom on the barrier as these couplings will be included in the coupled-channels calculations. 

In order to preserve the consistency with the microscopic calculations, it is necessary to compute the potential from the same EDF as in the HF and TDHF calculations. 
This is possible using the frozen HF technique \cite{sim12b}. 
Let us define an energy density $\mH[\xi(\vr)]$ such that $E[\rho]=\int d\vr  \mH[\xi(\vr)]$ where $\xi$ represents the set of local densities defined in section \ref{sec:skyrme}. 
We get the expression for the frozen potential 
\oeq
V(\vR)=\int \d\vr \sdf \mH[\xi_1(\vr)+\xi_2(\vr-\vR)] - E[\rho_1] -E[\rho_2],
\label{eq:frozen}
\ceq
where $\xi_{1,2}$ and $\rho_{1,2}$ are the local densities and one-body density matrices of  the 
HF ground-states of the fragments, respectively, and $\vR$ is the distance between their centers of mass. 
The isospin label $q$ has been omitted for simplicity. 

Equation (\ref{eq:frozen}) neglects the effect of the Pauli principle between pairs of nucleons belonging to different collision partners. 
In principle, one can include corrections to account for this effect using, e.g., the Thomas-Fermi approach \cite{den10a,den10b}. 
However, for light systems, the barrier is almost unaffected by the Pauli principle between the two reactants.
This is because, for such systems, the barrier is found at a relatively large distance between the nuclei where the overlap between the density distributions is small.
Of course, the inner part of the potential is associated to larger overlaps of the densities where the Pauli principle is expected to play a more important role. 
In the present work, we focus on the behaviour of fusion cross-sections near the barrier and, then, we neglect the Pauli principle in  the determination of the nucleus-nucleus potential. 
A better description of the inner part of the potential barrier should be considered to investigate, e.g., deep-sub-barrier fusion. 

\begin{figure}[!htb]
\includegraphics*[width=8.6cm]{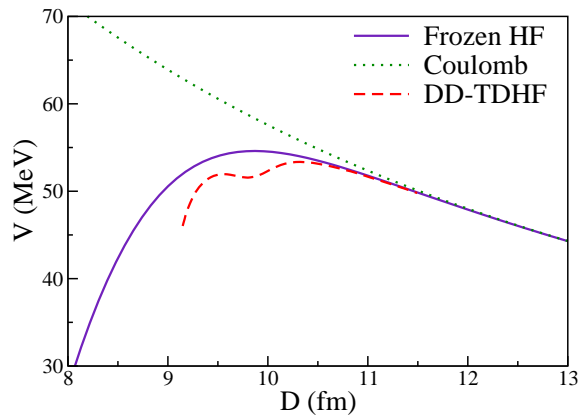}
\caption{(Color online) Frozen HF calculation of the nucleus-nucleus potential in the $^{40}$Ca+$^{40}$Ca system (solid line). The Coulomb part is shown with a dotted line. The DD-TDHF potential (dashed line) is from Ref.~\cite{was08}.}
\label{fig:Ca+Ca_frozen}
\end{figure}

The frozen HF potential of the $^{40}$Ca+$^{40}$Ca system is shown in Fig.~\ref{fig:Ca+Ca_frozen} (solid line). 
The barrier height is $V_B^{frozen}\simeq54.6$~MeV at a distance $D_B^{frozen}\simeq9.9$~fm. 
The DD-TDHF potential obtained from a TDHF calculation in Ref.~\cite{was08} at an energy close to the barrier is also reported. 
It is interesting to note that the dynamical effects included in the DD-TDHF potential lower the barrier by $\sim1.25$~MeV. 

\begin{figure}[!htb]
\includegraphics*[width=8.6cm]{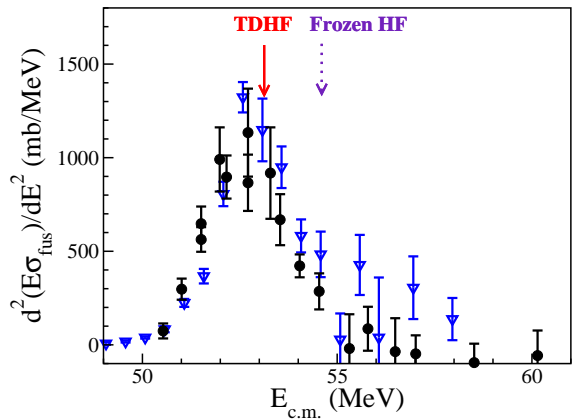}
\caption{(Color online) Experimental barrier distribution for $^{40}$Ca+$^{40}$Ca from data by Aljuwair {\it et al.} \cite{alj84} (filled circles) and from Montagnoli {\it et al.} \cite{mon12} (open triangles). The solid arrow indicates the TDHF fusion threshold (i.e., the expected position of the centroid of the barrier distribution), while the dashed one shows the position of the barrier from the frozen-HF technique.
The difference between frozen HF and TDHF barriers is due to collective couplings (see text).}
\label{fig:Ca+Ca_DB_exp}
\end{figure}

In addition, the experimental barrier distribution for this system is presented in Fig.~\ref{fig:Ca+Ca_DB_exp}. 
Its centroid is located $\sim2$~MeV below the frozen HF prediction. 
This indicates a possible effect of dynamical couplings (as expected due to vibrational couplings) not included in the frozen HF potential. 
However, these dynamical effects are included in the DD-TDHF potential at the mean-field level \cite{was08}.
As a result, the DD-TDHF barrier is in better agreement with the centroid of the experimental barrier distribution than the frozen HF barrier. 

Now, the question is: Can we recover the lowering of the barrier using standard coupled-channels calculations with the HF-frozen potential and couplings to collective vibrations?
Before addressing this question, however, we present a brief study of the dynamics of the $^{40}$Ca+$^{40}$Ca system at the barrier. 
As mentioned in the introduction, this step is not absolutely necessary to answer the above question. 
Nevertheless, these calculations may provide valuable information on the dynamics and could be used to select the most important collective modes affecting the reaction outcome. 
In particular, such calculations can be used to quantify the time during which the rearrangement of the density induced by the couplings and responsible for the dynamical modification of the potential occurs.

\section{Near-barrier TDHF calculations \label{sec:TDHF}}

\begin{figure}[!htb]
\includegraphics*[width=8.6cm]{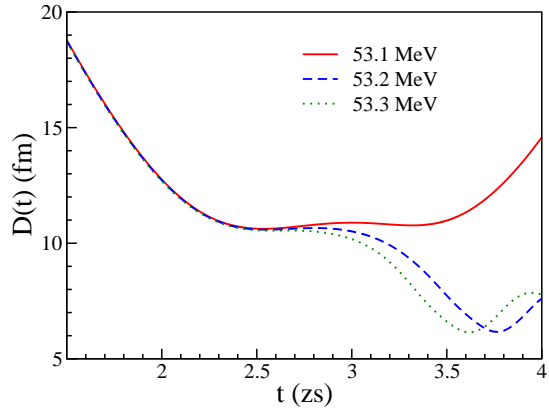}
\caption{(Color online) Time evolution of the distance between the fragments in  $^{40}$Ca+$^{40}$Ca central collisions.}
\label{fig:traj}
\end{figure}

TDHF calculations of $^{40}$Ca+$^{40}$Ca central collisions have been performed at energies around the barrier. 
Fig.~\ref{fig:traj} shows the time-evolution of the separation between the fragments for three selected energies.
The lowest energy leads to re-separation of the fragments, while the outcome of the reactions at higher energies is the formation of a compound nucleus. 

From these calculations, we deduce a TDHF energy threshold for this system $V_B^{TDHF}\simeq 53.15$~MeV (indicated by a solid arrow in Fig.~\ref{fig:Ca+Ca_DB_exp}). 
This energy is $\sim1.45$ MeV below the frozen HF barrier, in good agreement with the DD-TDHF calculations of Ref.~\cite{was08}. 
The latter point was expected as the DD-TDHF potential (see Fig.~\ref{fig:Ca+Ca_frozen}) is obtained from a TDHF evolution near the barrier. 
Note that the DD-TDHF barrier is still slightly higher than the present TDHF fusion threshold by $\sim0.2$~MeV.
This small difference can be attributed to the larger mesh size of $\Delta x=0.8$~fm used in Ref.~\cite{was08}. 
As discussed in section~\ref{sec:num}, the present calculations are performed with $\Delta x=0.6$~fm so that the violation of the Galilean invariance due to the grid affect the center of mass energy by less than 0.1~MeV. 
This allows us to investigate how the details of the trajectories shown in Fig.~\ref{fig:traj} are affected by a small change of the energy.
In particular, these calculations indicate that, at these near-barrier energies, the distance between the fragments is approximatly constant at $D\sim10-11$~fm for $\sim1$~zs. 
This is when the bifurcation between fusion and re-separation trajectories occurs.

\begin{figure}[!htb]
\includegraphics*[width=5cm]{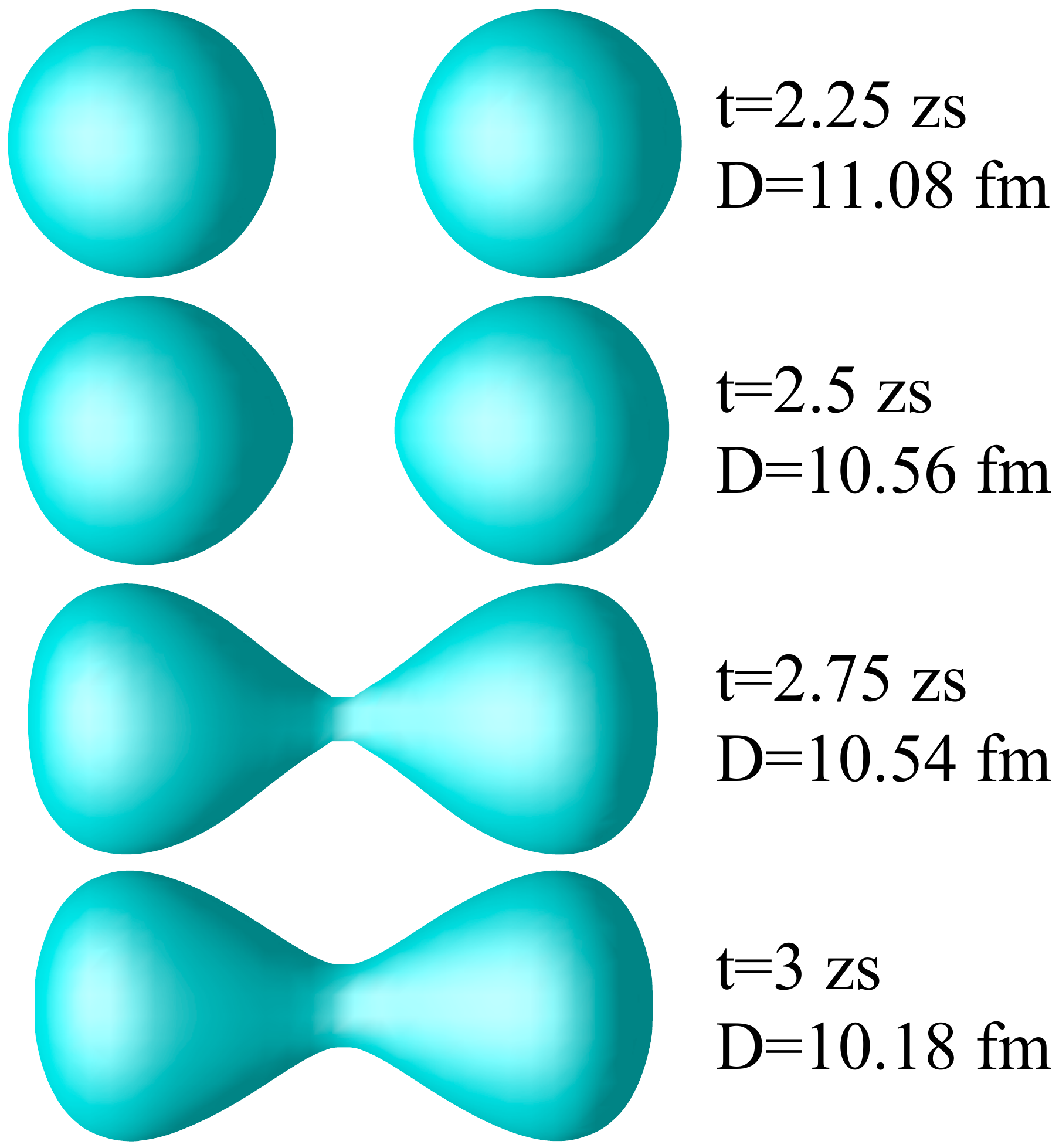}
\caption{(Color online) Isodensity surfaces with $\rho_0/2=0.08$~fm$^{-3}$ on a $^{40}$Ca+$^{40}$Ca central collision at $E_{c.m.}=53.3$~MeV.
The reference time $t=0$ corresponds to the initial condition of the TDHF calculation, i.e., with a separation distance between the nuclei $D_0=45.6$~fm.}
\label{fig:dens_CaCa}
\end{figure}

A deeper insight into the dynamics of the system during this critical period can be obtained from the density evolution. 
The latter is illustrated in Fig.~\ref{fig:dens_CaCa}.
It indicates that, at these distances, the system undergoes a rapid change of shape. 
The couplings induce then a rearrangement of the density within about $1$~zs.
It is interesting to note that, together with the formation of a neck between the fragments, the latter acquire a large octupole deformation.
This indicates, qualitatively, that the collective octupole modes in $^{40}$Ca seem to play an important role in the dynamics near the barrier. 
Note that other modes, such as quadrupole vibrations, might also affect the dynamics, although their effect cannot be directly deduced from  a simple observation of the density evolution in Fig.~\ref{fig:dens_CaCa}. 

In the next section, we present a method to investigate these collective vibrations with a TDHF code.

\section{Strength functions \label{sec:RPA}}

Modern TDHF codes have been used to perform real-time calculations of nuclear vibrations \cite{sim03,uma05,mar05,nak05,ste10,fra12}.
Indeed, the random phase approximation (RPA), which is a standard tool to investigate harmonic vibrations, can be obtained from the linearization of the TDHF equation.  
The strength functions (which are used to get the properties of the vibrational states in section~\ref{sec:param}) are then computed within the linear response theory. 

\subsection{Linear response theory \label{sec:linear-response}}

Let us consider a time-dependent perturbation 
\oeq
\oV(t)=\varepsilon f(t) \oQ
\label{eq:Voft}
\ceq
 applied on the ground state $|0\>$ of the system. 
The amplitude of the perturbation is quantified by $\varepsilon$, and its time-dependence by the function $f(t)$. 
The excitation operator $\oQ$ can be chosen, e.g., to be a multipole moment which, for $\la\ge2$, reads
\oeq
\oQ_{\la\mu}= \sum_{i=1}^A \hat{r}^\lambda \hat{Y}_{\la\mu}.
\ceq

This perturbation induces a time-evolution of the expectation value of the excitation operator $Q_{\la\mu}(t)=\<\Phi(t)|\oQ_{\la\mu}|\Phi(t)\>$. 
The strength function of the operator $\oQ_{\la\mu}$, defined as
\oeq
S_{\la\mu}(E)=\sum_{\nu}\ll\<\nu|\oQ_{\la\mu}|0\>\rl^2\delta(E-E_\nu+E_0),
\label{eq:def_strength}
\ceq
where $|\nu\>$ are the eigenstates of the Hamiltonian with energy $E_\nu$, is then obtained from \cite{ste04}
\oeq
{\tilde{f}(\omega)}S_{\la\mu}(\hb\omega)= \lim_{\varepsilon\rightarrow0}\frac{-1}{\pi\varepsilon} \mbox{Im}{\tilde{Q}_{\la\mu}(\omega)},
\label{eq:strength}
\ceq
where $\tilde{f}(\omega)$ and $\tilde{Q}_{\la\mu}(\omega)$ are the Fourier transforms of $f(t)$ and $Q_{\la\mu}(t)$, respectively. 

In calculating the strength function $S_{\la\mu}$ from Eq.~(\ref{eq:strength}), we then need:
\begin{itemize}
\item to specify the time-dependence of the excitation $f(t)$, or, equivalently, its Fourier transform $\tilde{f}(\om)$,
\item to consider an excitation intensity $\varepsilon$ to be sufficiently small to be in the linear regime, i.e., such that $Q_{\lambda\mu}(t)\propto\varepsilon$,
\item to determine  $Q_{\lambda\mu}(t)$ and its Fourier transform from a time-dependent model.
\end{itemize}
These points are described in more details in the following.

\subsection{TDHF numerical applications \label{sec:TDHF-vib}}

In the present work, we are interested in the effect of both low-lying and high energy collective vibrations on fusion. 
From a numerical point of view, the case of unbound states should be considered with care due to the echo generated by the reflection of emitted nucleons from the box boundaries \cite{rei06}. 
In principle, these reflections could have numerical effects on the entire strength distribution, including low-lying states. 
Note that emitted nucleons could be partially absorbed by applying an imaginary potential at the box boundaries \cite{nak05,rei06,par13}.
However, to avoid the increase of computational time associated with this technique, 
%In particular, low-lying collective states have to be treated properly as they are known to strongly affect the near-barrier reaction mechanisms \cite{mor94,ste95a,hag97}. 
another approach is considered to make sure that nucleons reflected on the box boundaries do not affect the transition amplitudes.
For this purpose, different techniques are used to compute the transition amplitudes of low-lying vibrations and those of giant resonances.

To avoid spurious effects of emitted nucleons on the strength distribution of bound states, we can adjust the time-dependence of the excitation operator in Eq.~(\ref{eq:Voft}) in such a way that unbound-states are not excited. 
Indeed, we see in the left-hand side of Eq.~(\ref{eq:strength}) that the strength distribution is multiplied by $\tilde{f}(\om)$ which can be chosen to be equal to 1 for $E\le \Omega$  and 0 for $E>\Omega$. 
This can be achieved by setting $f(t)=\frac{\sin(\Omega t/\hb)}{\pi t}$. 
$\Omega$ is chosen to be the particle emission threshold (neglecting tunneling), i.e., $\Omega\simeq\min\{S_n;S_p+B\}$ where $S_{n,p}$ are neutron and proton emission thresholds, and $B$ the Coulomb barrier for protons. 

\subsubsection{Application to low-lying vibrations in $^{40}$Ca}

As an example of application, the TDHF response to an octupole excitation $\varepsilon f(t)\oQ_{30}$ applied on the $^{40}$Ca HF ground-state has been computed in the linear regime over a finite time $T=15$~zs with $\Omega=12$~MeV. 
To account for the fact that $T$ is finite and to avoid spurious oscillations in the Fourier transform, $Q_{30}(t)$ is multiplied by a time-filtering function $\cos\frac{\pi t}{2T}$ \cite{rei06}, inducing an additional width of $\sim0.3$~MeV to the peaks. 

\begin{figure}[!htb]
\includegraphics*[width=8.6cm]{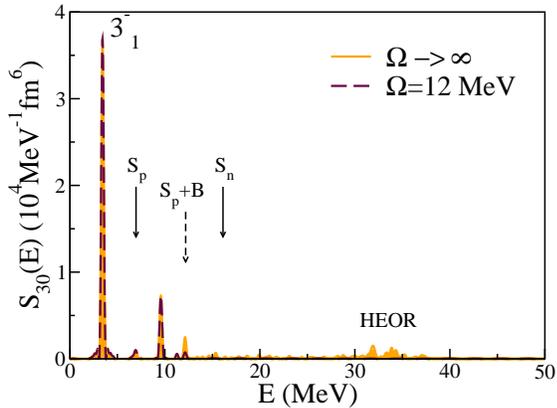}
\caption{(Color online) Strength function of the multipole moments in $^{40}$Ca.}
\label{fig:CaR3}
\end{figure}

The strength function computed from Eq.~(\ref{eq:strength}) is shown with a dashed line in Fig.~\ref{fig:CaR3}.
It is compared with the strength function obtained with $f(t)=\delta(t)$ (shaded spectrum), which is equivalent to $\Omega\rightarrow\infty$, i.e., all bound and unbound states are excited with no energy selection.  
In this case, states above 12~MeV are excited, in particular the high energy octupole resonance (HEOR) at $30-35$~MeV. 
As expected, the peaks above $12$~MeV are suppressed in the spectrum represented by the dashed line. 
Below this value, both spectra are identical indicating that particles emitted by the HEOR and reflected on the boundaries do not affect the strength of the low-lying states. 
This is, however, not necessarily the case with quadrupole vibrations as the GQR is much more collective than the HEOR.
Hence, a filtering function with finite  $\Omega$~MeV is applied consistently for the extraction of low-lying collective modes. 

As seen in Fig.~\ref{fig:CaR3}, the octupole strength distribution in $^{40}$Ca is dominated by the $3^-_1$ state at $E_{3^-_1}^{TDHF}=3.44$~MeV which is reasonably close to the experimental value $E_{3^-_1}^{Exp}=3.74$~MeV \cite{kib02}. 
The transition amplitude for this state, obtained by integrating the peak, is $|\<3^-_1|\oQ_{30}|0\>|\simeq113$~fm$^3$. 
The other states, including the HEOR, are at higher energy and have a smaller strength. 
The $3^{-}_1$ state is then expected to have a much stronger effect on near-barrier reaction mechanisms than the other octupole states.
Thus, the latter will be neglected in the coupled-channels calculations.

No other low-lying collective states were found in our calculations of quadrupole vibrations ($\la=2$) in $^{40}$Ca. 
This is consistent with the fact that, in this nucleus, the $2^+_1$ state does not exhibit a strong increase of collectivity as compared to the single-particle picture \cite{ram87}. 

\subsubsection{Low-lying states in $^{56}$Ni and role of magic numbers}

It is well known that all magic numbers induce an increase of the energy of the $2^+_1$ state \cite{ram87,spe89}. 
A reduction of the collectivity of the $2^+_1$ states is also observed for all magic numbers.
However, the effect is stronger for 20 than 28 \cite{ram87}. 
This is in qualitative agreement with the present TDHF calculations in which we found no low-lying collective $2^+$ state in $^{40}$Ca, while the $2^+_1$ state in $^{56}$Ni is found to be collective, with $E^{TDHF}_{2^+_1}\simeq3.02$~MeV and $\beta^{TDHF}_2=0.114$. 
Note that these predictions are in relatively good agreement with the experimental data $E^{Exp}_{2^+_1}\simeq2.7$~MeV and $\beta^{Exp}_2\simeq0.15-0.17$ \cite{yur04,jun11}. 

The energy of the $3^-_1$ states also increases in magic nuclei, but only for magic numbers 28 and above.
Indeed, no increase of $E_{3^-_1}$ is observed at $N,Z=8$ and 20 in the systematics of Refs~\cite{spe89,kib02}. 
This is why the $3^-_1$ is found at a rather low energy of 3.44~MeV in $^{40}$Ca, but at a relatively high energy $E_{3^-_1}\simeq9.64$~MeV in $^{56}$Ni, with a deformation parameter $\beta_3\simeq0.127$. 
In fact, this state is found to be (quasi-)bound only by the Coulomb barrier. 

We see that the low-lying vibrational spectra varies for magic numbers 20  and 28. 
These differences are due to the spin-orbit interaction which is responsible for the magic number 28.
The above discussion shows the importance of a dynamical model which accounts for the specificities of the vibrational spectra in each nuclei. 
This is possible in modern TDHF calculations thanks to a fully microscopic treatment of both the structure and the reaction dynamics using modern energy density functionals including spin-orbit interactions. 

\subsubsection{Case of giant resonances \label{sec:GR}}

The giant monopole resonance (GMR, $\la=0$), the isovector giant dipole resonance (IVGDR, $\la=1$), and the giant quadrupole resonance (GQR, $\la=2$) are known to account for a large part (if not all) of their associated energy weighted sum rule \cite{har01}. 
Due to its isovector nature, the IVGDR modifies essentially the difference between proton and neutron densities, while, their sum is almost unchanged. 
It is then reasonable to assume that the IVGDR may affect only the Coulomb part of the nucleus-nucleus potential. 
For light systems, however, $Z_1Z_2$ is small and the Coulomb coupling to the IVGDR can be neglected in a first approximation. 

The GMR and GQR are both isoscalar modes and might modify the nuclear attraction between the reactants. 
However, the GMR is usually located at a higher energy than the GQR and, then, should be less coupled to the relative motion. 
%In addition, coupled-channels codes usually incorporate couplings to $\la\ge2$ modes. 
An estimate of the effect of couplings to giant resonances on fusion will then be obtained by focussing on the GQR, while neglecting couplings to the other resonances. 

The details of the strength distribution in the giant resonance region, in particular their fragmentation, are known to be sensitive to the spurious reflection of particles on the numerical box boundaries \cite{rei06}. 
For the purpose of this work, however, it is sufficient to assume that the giant resonances are concentrated in one single peak. 
With this assumption, the characteristics of the GQR can be extracted from the first minimum of $Q_{20}(t)$ following the perturbation $\varepsilon\delta(t)\oQ_{20}$  in the linear regime using \cite{sim03,sim09}
\oeq
Q_{20}(t)\simeq\frac{-2\varepsilon}{\hb}\ll\<GQR|\oQ_{20}|0\>\rl^2\sin(E_{GQR}t/\hb).
\ceq
As a result, we get for $^{40}$Ca an energy $E_{GQR}^{TDHF}\simeq18.1$~MeV and a transition amplitude $\ll\<GQR|\oQ_{20}|0\>\rl\simeq20.53$~fm$^{2}$. 
For the $^{56}$Ni nucleus, we get $E_{GQR}^{TDHF}=16.8$~MeV and $\beta_2=0.116$.

\section{Deformation parameters of vibrational states \label{sec:param}}

The deformation parameter $\beta_{\la}^{(\nu)}$ of a vibrational state $|\nu\>$ with angular momentum $\la$ is a critical input to describe its coupling to relative motion in coupled-channels calculations.
Here, we show how to extract these parameters directly from the strength distribution. 

Consider a small excitation potential $\varepsilon\delta(t)\oQ_{\la0}$ with $\la\ge2$. 
In the first order in $\varepsilon$, the wave-function reads (see, e.g., Refs.~\cite{sim03,sim09})
\oeq
|\Phi(t>0)\> \simeq e^{-iE_0t/\hb}\( |0\> -\frac{i\varepsilon}{\hb} \sum_\nu q_\nu e^{-i\omega_\nu t} |\nu\> \), \label{eq:phi}
\ceq
where $q_\nu=\<\nu|\oQ_{\la0}|0\>$ is the transition amplitude between the ground-state $|0\>$ and the Hamiltonian eigenstate $|\nu\>$ with energy $E_\nu$ and angular momentum $\la_\nu=\la$. 
We have introduced $\omega_\nu=(E_\nu-E_0)/\hb$. 
For simplicity, we assume that $q_\nu$ is real.

Using Eq.~(\ref{eq:phi}) and the observable $\oR(\theta\phi)$ measuring the distance to the surface of the nucleus, defined in Eq.~(\ref{eq:oR}), and 
noting that $\<\nu|\oad_{\la0}|0\>=\delta_{\la\la_\nu}\sqrt{2\la+1}$ we get
 \oeq
 \<\delta\oR(\theta\phi)\> = \frac{-2\varepsilon}{\hb}R_0Y_{\la0}(\theta) \sum_\nu \be_{\la}^{(\nu)} q_\nu\sin{\omega_\nu t},
\ceq
where $\delta\oR=\oR-R_0$. 
Assuming a constant density with a sharp surface, this surface variation can be related to the expectation value of the multipole moment as
\oeq
Q_{\la0}(t)=\frac{-3\varepsilon}{2\pi\hb}  A R_0^\la \sum_\nu \be_{\la}^{(\nu)} q_\nu \sin \omega_\nu t.
\ceq
Using Eq.~(\ref{eq:strength}), the strength function can be expressed as
\oeqn
S_{\la0}(E) &=& \frac{-1}{\pi\varepsilon} \int_0^\infty \stb dt\stf Q_{\la0}(t)\sin (Et/\hb)  \nonumber \\
&=& \frac{3}{4\pi} A R_0^\la \sum_\nu \be_\la^{(\nu)} q_\nu \delta(E-E_\nu+E_0).
\ceqn
Identifying with Eq.~(\ref{eq:def_strength}), we get
\oeq
\be_\la^{(\nu)} = \frac{4\pi q_\nu}{3 A R_0^\la}.
\label{eq:beta1}
\ceq

Usually, the deformation parameters are determined from the experimental reduced electric transition probability $B(E\la;0^+_1\rightarrow\nu)$ data using \cite{hag12}
\oeq
\be_\la^{(\nu)} = \frac{4\pi }{3 Z R_0^\la} \sqrt{\frac{B(E\la;0^+_1\rightarrow\nu)}{e^2}}.
\label{eq:beta2}
\ceq
 Equations (\ref{eq:beta1}) and (\ref{eq:beta2}) are equivalent if one assumes an exact proton-neutron symmetry. 
 This assumption should be valid for light $N=Z$ nuclei like the one studied here. 
 However, for heavier nuclei, corrections might have to be considered on the experimental deformation parameter extracted from Eq.~(\ref{eq:beta2}) due to differences between proton and neutron densities. 

\begin{figure}[!htb]
\includegraphics*[width=8.6cm]{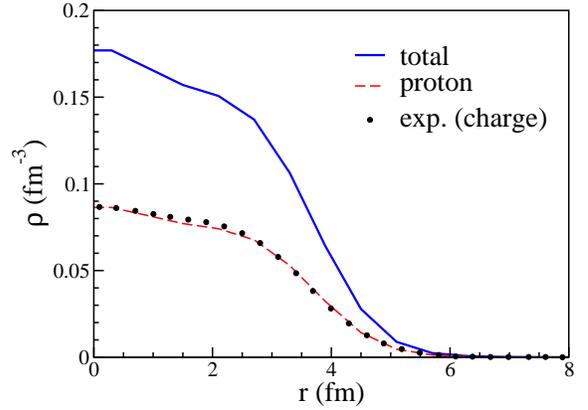}
\caption{(Color online) Radial density in $^{40}$Ca. The total (solid line) and proton (dashed line) densities are those of the HF ground-state. 
The circles are the experimental charge density \cite{neg82}. }
\label{fig:rho}
\end{figure}

It is important to note that the deformation parameter is quite sensitive to the radius of the nucleus, in particular for high multipolarities, as it is proportional to $R_0^{-\la}$. 
The HF proton density shown in Fig.~\ref{fig:rho} with a dashed line for $^{40}$Ca is in fact in excellent agreement with the experimental charge density~\cite{neg82}.
Note that the HF calculations predict a neutron density (not shown in Fig.~\ref{fig:rho} for clarity) which is very similar to the proton one, supporting the validity of the proton-neutron symmetry in this nucleus. 
The nuclear radius $R_0$ is then determined from the total HF ground-state density shown by the solid line in Fig.~\ref{fig:rho}. 
Using $\rho_0/2=0.08$~fm$^{-3}$, we get $R_0\simeq3.67$~fm. 
This corresponds to a radius parameter $r_0=R_0A^{-\frac{1}{3}}\simeq1.07$~fm. 

Using Eq.~(\ref{eq:beta1}) and the strength distribution in Fig.~\ref{fig:CaR3}, we then obtain a deformation parameter of the $3^-_1$ state in $^{40}$Ca of $\beta_3\simeq0.24$. 
This value is smaller than the experimental deformation parameter $\beta_3\simeq0.3-0.4$ obtained from angular distributions of inelastically scattered protons \cite{kib02}. 
However, it should be emphasized that it is obtained from purely microscopic calculations, as the only input parameters are those of the Skyrme EDF. 

The deformation parameter associated with the GQR can also be extracted from Eq.~(\ref{eq:beta1}).
As a result we get $\beta_{2}\simeq0.16$. 
The fact that the deformation parameter is smaller, and the energy higher, than the low-lying octupole state implies that the GQR is only expected to provide a small correction to the barrier distribution and that the main effects will be obtained from the coupling to the $3^{-}_1$ state. 
This will be confirmed with the coupled-channels calculations presented in the next section.

\section{Coupled-channels calculations \label{sec:cc}}

The coupled-channels calculations are performed with the \textsc{ccfull} code \cite{hag99}. 
The bare nucleus-nucleus potential obtained with the frozen HF model [Eq.~(\ref{eq:frozen})] is used. 
However, the nucleus-nucleus potential is assumed to have a Woods-Saxon form in the \textsc{ccfull} code. 
The nuclear part of the frozen HF potential in Fig.~\ref{fig:Ca+Ca_frozen} has then been fitted with a Woods-Saxon function [Eq.~(\ref{eq:WS})]
The fit is performed in the region $R>8.3$~fm to allow an accurate reproduction of the barrier height and position. 
The parameters of the fit are given in table \ref{tab:WS}. 
Note that, in the \textsc{ccfull} code, this potential is also used to evaluate the coupling matrix elements. 
In principle, one may extract the coupling form factors (that is, the off-diagonal part of the potential) directly from TDHF calculations. 
This extension of the present method will be the subject future investigations. 
  
 \begin{table}
 \caption{\label{tab:WS}
Woods-Saxon parametrization of the nucleus-nucleus potentials.}
 \begin{ruledtabular}
 \begin{tabular}{cccc}
System & $V_0$ (MeV) & $r_0$ (fm) & $a$ (fm)\\
\hline
$^{40}$Ca+$^{40}$Ca        & 98.7 & 1.146 & 0.629 \\
$^{56}$Ni+$^{56}$Ni        & 132.5 & 1.103 & 0.631 \\
 \end{tabular}
 \end{ruledtabular}
 \end{table}

The coupled-channels calculations of the $^{40}$Ca+$^{40}$Ca reaction include couplings to the $3^-_1$ and GQR states.
Note that the coupling to higher energy and less collective octupole states (neglected here) could also affect the fusion cross-sections as shown in Ref.~\cite{row10}.
The energy of the vibrational states and the coupling strengths (deformation parameters) have been computed with the \textsc{tdhf3d} code using the linear response theory (see table~\ref{tab:coup}).

  \begin{table}
 \caption{\label{tab:coup}
Energy and deformation parameter of the collective vibrational states from TDHF calculations.}
 \begin{ruledtabular}
 \begin{tabular}{cccc}
Nucleus & State & $E_\nu$  & $\beta_\nu$ \\
\hline
$^{40}$Ca 	& $3^-_1$ 	& 3.44	& 0.240 \\
		        & GQR 		& 18.1 	& 0.160 \\
$^{56}$Ni         & $2^+_1$ 	& 3.02 	& 0.114 \\
			& $3^-_1$ 	& 9.64 	& 0.127 \\
		        & GQR 		& 16.8 	& 0.116 \\
 \end{tabular}
 \end{ruledtabular}
 \end{table}

\begin{figure}[!htb]
\includegraphics*[width=8cm]{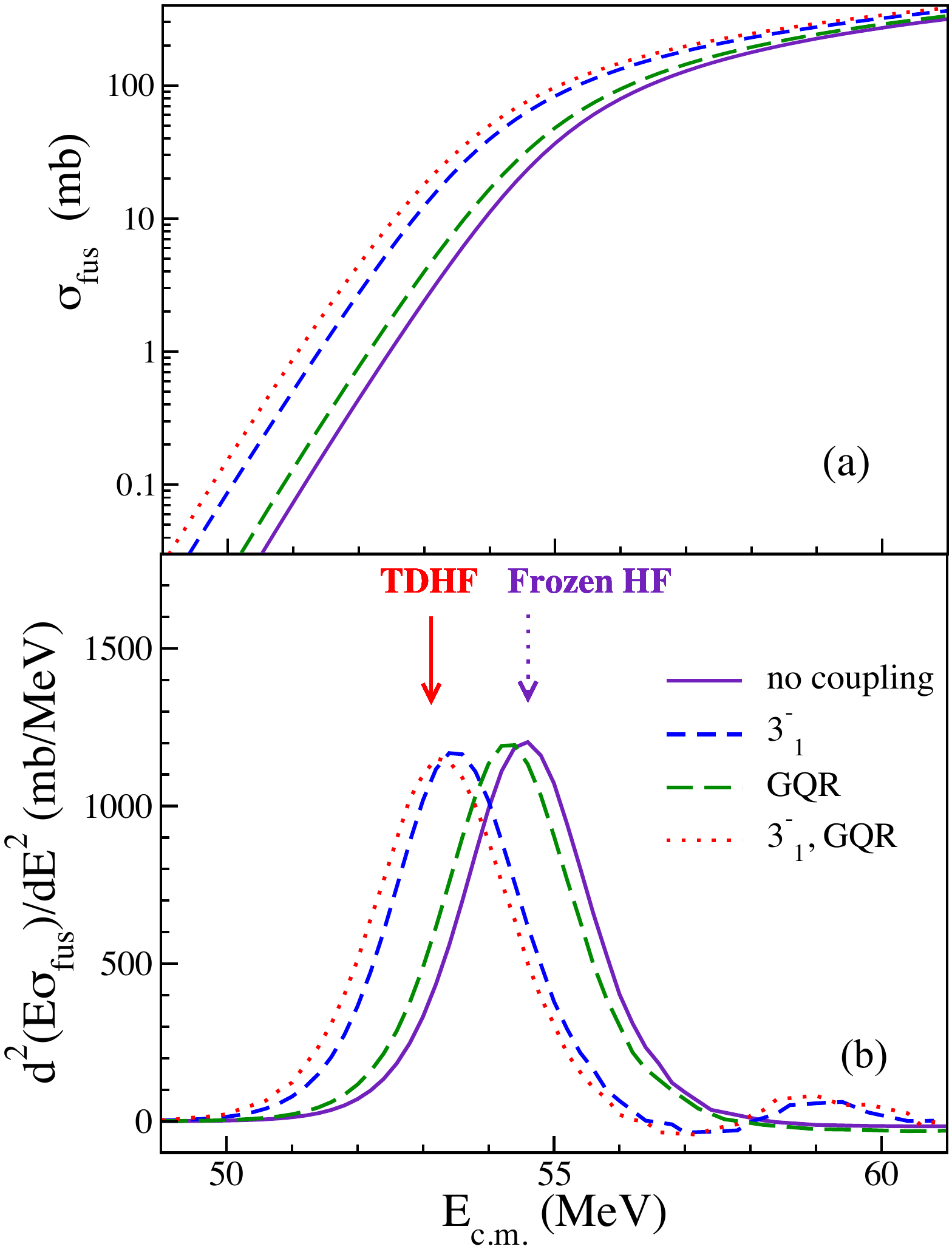}
\caption{(Color online) Coupled-channels calculations of (a) fusion cross-sections and (b) barrier distributions in $^{40}$Ca+$^{40}$Ca. The solid arrow indicates the TDHF fusion threshold, while the dashed one shows the position of the barrier from the frozen-HF technique.}
\label{fig:Ca+Ca_DB_CC}
\end{figure}

In principle, the \textsc{ccfull} code allows for the inclusion of two vibrational modes in the target and one in the projectile \cite{hag99}.
However, for symmetric systems, the mutual excitation of one-phonon states in each collision partner can be included by considering only one phonon with a coupling constant $\sqrt{2}\beta_\lambda$ \cite{esb87}. 
Using this, we are then able to include up to three vibrational modes in both symmetric collision partners.  

The resulting fusion cross-sections are shown in Fig.~\ref{fig:Ca+Ca_DB_CC}(a) for the $^{40}$Ca+$^{40}$Ca reaction. 
The barrier distributions are computed from these cross-sections using a three-point formula with $\Delta E=0.2$~MeV. 
The barrier distribution is shown in Fig.~\ref{fig:Ca+Ca_DB_CC}(b). 
The variation of the centroids of these distributions are given in table~\ref{tab:centroids}.
The coupling to the ${3}^-_1$ state has a strong effect on the barrier distribution.
Indeed, it induces a lowering of the centroid by $\sim1.1$~MeV, while the coupling to the GQR renormalizes the potential by only $\sim-0.3$~MeV. 
The inclusion of the second phonon of the octupole low-lying vibrational mode (not shown in Fig.~\ref{fig:Ca+Ca_DB_CC} for clarity) has a negligible effect on the barrier centroid, as it reduces it by less than $40$~keV compared to the one-phonon case. 
We see in Fig.~\ref{fig:Ca+Ca_DB_CC} that these couplings explain well the global lowering of the fusion threshold obtained with TDHF in comparison with the frozen HF barrier. 
As a result, they increase the sub-barrier fusion cross-section by more than one order of magnitude.
Note that other collective vibrations not considered explicitly here, such as the GDR and the GMR, will reduce further the barrier centroid. 
However, their effect is expected to be smaller than the GQR, as discussed in section~\ref{sec:GR}. 

 \begin{table}
 \caption{\label{tab:centroids}
Difference between the centroids of the barrier distributions and the frozen HF barrier.}
 \begin{ruledtabular}
 \begin{tabular}{cccccc}
System 				& $2^+_1$ 	& $3^-_1$		& GQR	& $\{2^+_1$, $3^-_1$, GQR$\}$ 	& TDHF \\
\hline
$^{40}$Ca+$^{40}$Ca 	& 	 		& $-1.1$ 		& $-0.4$ 	& $-1.3$					& $-1.45$  \\
$^{56}$Ni+$^{56}$Ni 	& $-0.6$		& $-0.8$		& $-0.5$	& $-1.7$ 					& $-2.5$	\\
 \end{tabular}
 \end{ruledtabular}
 \end{table}

\section{Comparison with the $^{56}$Ni+$^{56}$Ni system \label{sec:nini}}

The $^{56}$Ni isotope is not stable, with a half-life of $\sim6$ days. 
This makes the experimental study of the $^{56}$Ni+$^{56}$Ni system with present accelerator facilities almost impossible. 
However, experimental barrier distributions have been measured for the $^{58,60}$Ni+$^{60}$Ni systems \cite{ste95a,rod06}. 
The shape of the experimental barrier distributions were found to be similar for these two systems, and could be  explained by invoking low-lying quadrupole vibrations alone. 
However, the energy of the $2^+_1$ states being very low in $^{58,60}$Ni (around 1.4~MeV, i.e., approximatively half the energy of the $2^+_1$ state in the doubly-magic $^{56}$Ni isotope), up to four phonons were needed to reproduce the shape of the barrier distribution in these systems. 

\begin{figure}[!htb]
\includegraphics*[width=8.6cm]{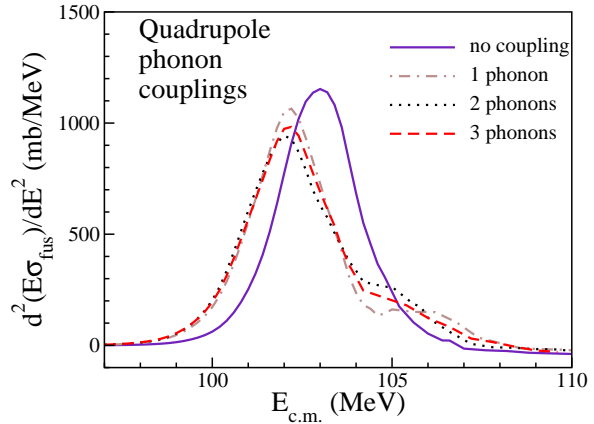}
\caption{(Color online) Coupled-channels calculations of barrier distributions in $^{56}$Ni+$^{56}$Ni with quadrupole phonons only.
The $n$-phonon states in each collision partner are included explicitly, i.e., without using the effective coupling $\sqrt{2}\beta_2$ for symmetric systems. }
\label{fig:DB_NiNi_quad}
\end{figure}

The situation is different in the $^{56}$Ni+$^{56}$Ni because the $2^{+}_1$ state is at higher energy in $^{56}$Ni, calculated to be $E_{2^+_1}^{TDHF}\simeq3.02$~MeV with the SLy4$d$ parametrization. 
Coupled-channel calculations have been performed to investigate the role of the number of quadrupole phonons on the barrier distributions. 
The parameters of the Woods-Saxon fit of the frozen HF potential are given in table~\ref{tab:WS} and the parameters of the quadrupole coupling in table~\ref{tab:coup}.
The barrier distributions are presented in Fig. \ref{fig:DB_NiNi_quad} for couplings to one, two, and three phonons. 
The inclusion of more than one quadrupole phonon in this system has little effect on the centroid of the barrier distribution.
Indeed, the centroid varies by less than 70~keV when one increases the number of phonons from one to three. 
In fact, adding more phonons essentially smears out the barrier distribution at higher energies. 

\begin{figure}[!htb]
\includegraphics*[width=8.6cm]{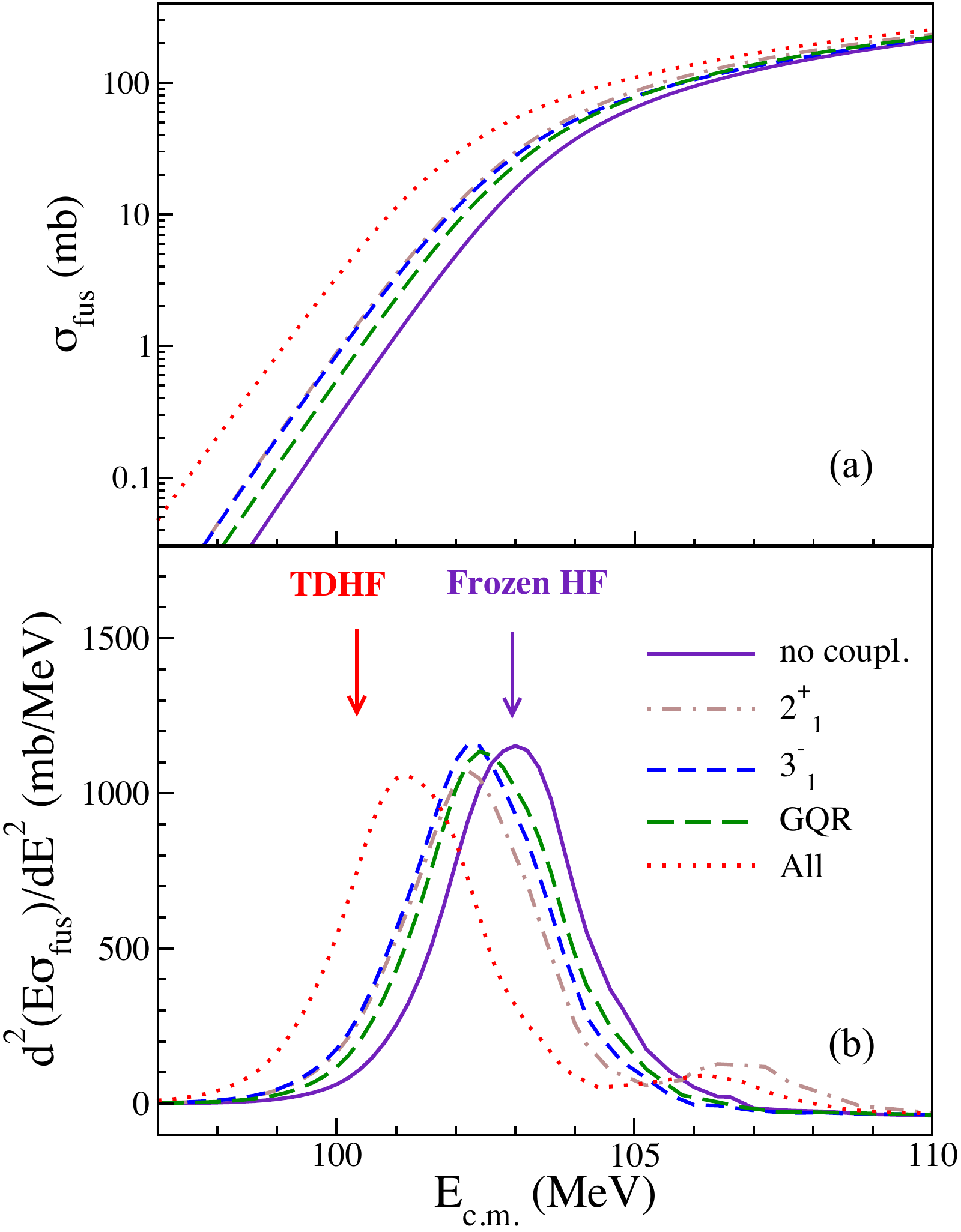}
\caption{(Color online) Coupled-channels calculations of (a) fusion cross-sections and (b) barrier distributions in $^{56}$Ni+$^{56}$Ni.
Only one-phonon states are considered.
The effective coupling $\sqrt{2}\beta_\lambda$ for symmetric systems is used. The solid arrow indicates the TDHF fusion threshold, while the dashed one shows the position of the barrier from the frozen-HF technique.}
\label{fig:Ni+Ni_DB}
\end{figure}

Coupled-channel calculations have been also performed to investigate the effect of the different vibrational modes on the barrier distribution in $^{56}$Ni+$^{56}$Ni. 
The fusion cross-sections and resulting barrier-distributions are presented in Figs.~\ref{fig:Ni+Ni_DB}(a) and~\ref{fig:Ni+Ni_DB}(b), respectively,
The barrier centroids are reported in table~\ref{tab:centroids}. 
Although the low-lying quadrupole vibration is the only one to affect the shape of the barrier distribution, each phonon, i.e., the $2^+_1$, the $3^-_1$ and the GQR state, lowers the centroid by a similar energy of $0.5$ to $0.8$~MeV.

Comparing the TDHF fusion threshold with the barrier distribution when all these couplings are included (see table~\ref{tab:centroids}), we see that the TDHF prediction is lying $\sim0.8$~MeV lower than the barrier centroid calculated with the coupled-channels approach. 
This indicates that other states should probably be included. 
For instance, other high-lying modes like the GDR might play a more significant role than in the $^{40}$Ca+$^{40}$Ca case due to the larger value of $Z_1Z_2$. 
The coupling to $1^-$ isovector states should  be included in coupled-channels codes to test their effect on barrier distributions. 

\begin{figure}[!htb]
\includegraphics*[width=5cm]{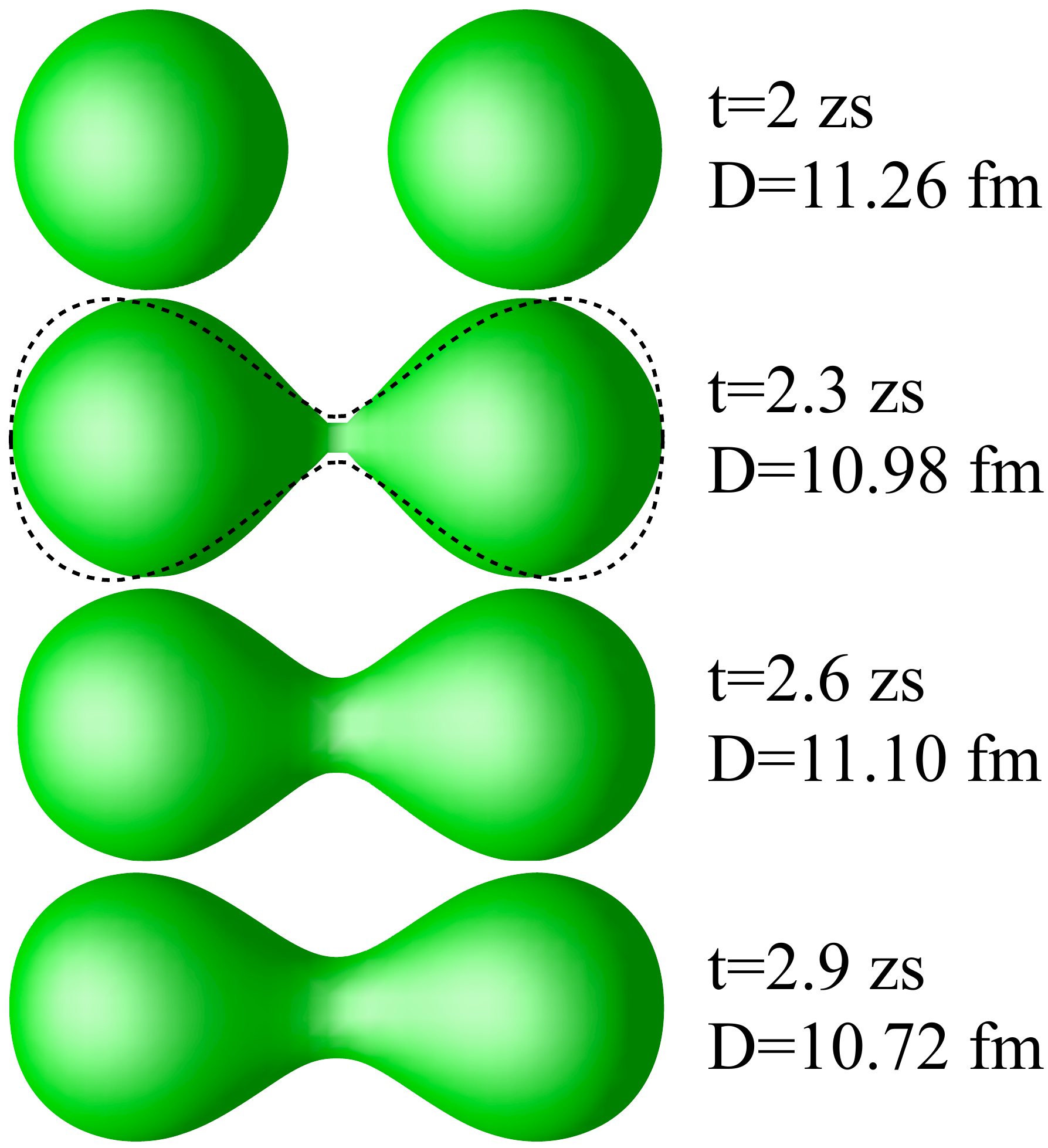}
\caption{(Color online) Isodensity surfaces with $\rho_0/2=0.08$~fm$^{-3}$ in a $^{56}$Ni+$^{56}$Ni central collision at $E_{c.m.}=100.5$~MeV. The dashed contour shows the same isodensity [with a magnification factor $R_0(^{56}$Ni$)/R_0(^{40}$Ca$)$] in $^{40}$Ca+$^{40}$Ca central collision at $E_{c.m.}=53.3$~MeV at the time when the neck is formed (see Fig.~\ref{fig:dens_CaCa}).}
\label{fig:dens_NiNi}
\end{figure}

Finally, we conclude this section by studying the evolution of the shape of the system near the barrier in Fig.~\ref{fig:dens_NiNi}. 
As in the $^{40}$Ca+$^{40}$Ca case, we see that the two fragments spend a relatively long time (about 1~zs in the $^{56}$Ni+$^{56}$Ni reaction  just above the barrier at $E_{c.m.}=100.5$~MeV) at an almost fixed distance, during which their shapes undergo large deformations. 
However, comparing with the $^{40}$Ca+$^{40}$Ca system when the neck is forming (see dashed contour in Fig.~\ref{fig:dens_NiNi}), we see that the two systems encounter rather different shapes. 
At variance with the $^{40}$Ca fragments, the $^{56}$Ni collision partners do not acquire a strong octupole deformation. 
This is consistent with the smaller effect of the $3^-_1$ state on $^{56}$Ni+$^{56}$Ni fusion.   

\section{Conclusions}

A technique is introduced to investigate the effect of collective vibrations on  fusion cross-sections where the only inputs are the choice of the collective phonons and the parameters of the Skyrme energy density functional. 
The coupled-channels model is used with potential and coupling parameters extracted from microscopic quantum calculations. 
The bare nucleus-nucleus potential is computed from the frozen Hartree-Fock technique, while the energy and deformation of the collective vibrational states are calculated with a time-dependent Hartree-Fock code using the linear response theory. 
The same TDHF code is used to investigate the dynamics of the density at near-barrier energies and to determine the fusion threshold where all the dynamical couplings are included to all orders in the mean-field approximation. 

The near-barrier fusion of two $^{40}$Ca and of two $^{56}$Ni nuclei has been investigated.
The TDHF fusion threshold, which automatically incorporates effects of collective couplings at the mean-field level, is considerably lower than the bare potential barrier in these systems, as would be expected. 
Low-lying collective vibrations, such as the $3^-_1$ state in $^{40}$Ca and the $2^+_1$ state in $^{56}$Ni, affect both the shape of the barrier distribution and the position of the main barrier. 
While in the $^{40}$Ca+$^{40}$Ca reaction the inclusion of the coupling to the $3^-_1$ state accounts for almost all the dynamical lowering of the barrier, the situation is more complex in the $^{56}$Ni+$^{56}$Ni system. 
Indeed, in the latter case,  states at higher energies, such as the giant quadrupole resonance or the octupole mode, induce a renormalization of the potential of the same order as the low-lying $2^+_1$ state. 
In fact, a comparison with the TDHF fusion threshold indicates that other modes, not included in the present description (e.g., the giant dipole resonance), might contribute to the  lowering of the barrier as well. 

An interesting aspect of the method is the consistency between each step of the description as they all use the same Skyrme EDF. 
In particular, the calculations do not depend on a particular choice of nucleus-nucleus potential parametrization, neither do they rely on experimental data. 
This technique can then be applied to systems where little is known, such as reactions involving exotic nuclei. 

It is important to note, however, that a perfect agreement  with experimental data should not always be expected as both the TDHF and the coupled-channels calculations have underlying approximations. 
For instance, the TDHF approach neglects pairing correlations and other residual interactions. 
Nevertheless, the technique is quite general and could benefit from improvements in one or the other models.
As an example, recent beyond TDHF codes including pairing correlations \cite{ave08,eba10,ste11,sca13} could be used as well.  

The calculations could also be affected by the choice of the parametrization of the Skyrme functional. 
Although most Skyrme functionals have been tested on giant resonance properties, few calculations have been made for low-lying collective states. 
It is shown in the present work that the expected behavior of the quadrupole and octupole modes with the magic numbers 20 and 28 is qualitatively well reproduced with the present functional. 
Nevertheless, more  tests across the nuclear chart should be performed. 
Systematic calculations of low-lying collective vibrations with different Skyrme functionals will be the subject of future works. 

Finally, the present paper focuses on the role of couplings to vibrational states, hence our choice of spherical symmetric systems to avoid effects from static deformations and from transfer. 
The present technique needs to be tested with deformed systems and in the more challenging case of couplings to transfer channels in asymmetric systems. 
Improvements of the method are also considered, such as extracting directly the coupling form factors from TDHF calculations, and taking into account anharmonic effects in the vibrational spectra by going beyond the linear response regime.
Deep sub-barrier fusion could also be investigated with an improvement of the description of the nucleus-nucleus potential at small relative distances between the collision partners. 

\begin{acknowledgments}

N. Rowley is thanked for useful discussions at the early stage of this work. 
We are grateful to K. Hagino for useful comments after a careful reading of the manuscript. 
Discussions with S. Umar and J. Maruhn are also acknowledged.
This work has been supported by the Australian Research Council under the FT120100760, FL110100098 and DP1094947 grants.
The calculations have been performed on the NCI National Facility in Canberra, Australia, which is supported by the Australian Commonwealth Government.

\end{acknowledgments}

\bibliography{biblio}

\end{document}